\documentclass[10pt,a4paper]{article}
\usepackage{RR}
\usepackage{hyperref}

\usepackage{graphicx,amsmath,amsfonts,psfrag}
\usepackage[normalem]{ulem}

\newcommand {\defin} {\stackrel{\Delta}{=}}

\def\junk#1{}

\RRdate{August 2006}
\RRauthor{Dinesh Kumar 
  \and 
Eitan Altman
  \and 
Jean-Marc Kelif
}
\authorhead{Dinesh Kumar, Eitan Altman \& Jean-Marc Kelif}
\RRtitle{Association d'Utilisateur-Réseau dans une Cellule Hybride de WLAN-UMTS : Optimalité Globale \& Individuelle}
\RRetitle{User-Network Association in a WLAN-UMTS Hybrid Cell: Global \& Individual Optimality}
\titlehead{User-Network Association in a WLAN-UMTS Hybrid Cell}
\RRnote{Dinesh Kumar and Eitan Altman are at INRIA, BP 93, 06902 Sophia Antipolis, France.
\\
(\{dkumar, altman\}@sophia.inria.fr)}
\RRnote{Jean-Marc Kelif is at France Telecom R\&D, 92794 Issy les Moulineaux, France.
\\
(jeanmarc.kelif@orange-ft.com)}

\RRnote{The authors of this report are thankful to France Telecom R\&D who financially
supported this work under the external research contract (CRE) no. 46130622.}

\RRresume{Nous étudions l'association optimale utilisateur-réseau dans une {\em cellule hybride} 802.11 WLAN et 3G-UMTS. En supposant que l'attribution de ressource sature le lien descendant des réseaux WLAN et UMTS et que les mobiles se situent tous à une même position 
{\em moyenne} dans la cellule hybride et appartiennent à la même classe de qualité de service, nous formulons le problème
selon deux approches différentes : optimalité globale et individuelle. L'association globalement optimale est formulée comme un problème de décision de routage SMDP (Semi Markov Decision Process) dans lequel les récompenses comportent une composante financière de gain et une composante de débit global de réseau. Les équations de programmation dynamique correspondantes sont résolues en utilisant la méthode d'itération des valeurs et la politique optimale stationnaire est alors obtenue avec une courbe de commutation ni convexe ni concave. Nous constatons que pour les cas analogues de réseau homogènes, les courbes de commutation sont symetriques et de type seuil. L'optimalité individuelle est étudiée dans un cadre de jeu dynamique non-coopératif en considérant le temps de service moyen d'un mobile comme critère de coût pour la décision. Nous montrons dans le cas de l'optimalité individuelle dans une cellule hybride de WLAN-UMTS, que la courbe de politique de seuil est une fonction décroissante par palier du taux d'arrivée de Poisson des mobiles.}

\RRabstract{We study optimal user-network association in an integrated 802.11
WLAN and 3G-UMTS {\em hybrid cell}. Assuming saturated resource
allocation on the downlink of WLAN and UMTS networks and a single
QoS class of mobiles arriving at an {\em average} location in the
hybrid cell, we formulate the problem with two different approaches:
Global and Individual optimality. The Globally optimal association
is formulated as an SMDP (Semi Markov Decision Process) connection
routing decision problem where rewards comprise a financial gain
component and an aggregate network throughput component. The
corresponding Dynamic Programming equations are solved using Value
Iteration method and a stationary optimal policy with neither convex
nor concave type switching curve structure is obtained. Threshold
type and symmetric switching curves are observed for the analogous
homogenous network cases. The Individual optimality is studied under
a non-cooperative dynamic game framework with expected service time
of a mobile as the decision cost criteria. It is shown that
individual optimality in a WLAN-UMTS hybrid cell, results in a
threshold policy curve of descending staircase form with increasing
Poisson arrival rate of mobiles.}
\RRmotcle{hybride, hétérogène, WLAN, UMTS, MDP, optimisation, commande, jeu non-coopératif}
\RRkeyword{hybrid, heterogeneous, WLAN, UMTS, MDP, optimization, control, non-cooperative game}
\RRprojets{Maestro}
\RRtheme{\THCom} 
\URSophia 
\begin{document}
\makeRR   
\section{Introduction}
As 802.11 WLANs and 3G-UMTS cellular coverage networks are being
widely deployed, network operators are seeking to offer seamless and
ubiquitous connectivity for high-speed wireless broadband services,
through integrated WLAN and UMTS hybrid networks. For efficient
performance of such an hybrid network, one of the core decision
problems that a network operator is faced with is that of optimal
user-network association, or load balancing by optimally routing an
arriving mobile user's connection to one of the two constituent
networks. We study this decision problem under a simplifying
assumption of saturated downlink resource allocation in the lone
WLAN and UMTS cells. To be more specific, consider a hybrid network
comprising two independent 802.11 WLAN and 3G-UMTS networks, that
offers connectivity to mobile users arriving in the combined
coverage area of these two networks. By independent we mean that
transmission activity in one network does not create interference in
the other. Our goal in this paper is to study the dynamics of
optimal user-network association in such a WLAN-UMTS hybrid network.
We provide two different and alternate modeling approaches that
differ according to who takes the {\it association} or {\it
connection} decision and what his/her objectives are. In particular,
we study two different dynamic models and the choice of each model
depends on whether the optimal objective criteria can be represented
as a global utility such as the aggregate network throughput, or an
individual cost such as the service time of a mobile user. We
concentrate only on streaming and interactive data transfers.
Moreover, we consider only a single QoS class of mobiles arriving at
an {\em average} location in the hybrid network and these mobiles
have to be admitted to one of the two WLAN or UMTS networks. Note
that we do not propose a full fledged cell-load or interference
based connection admission control (CAC) policy in this paper. We
instead assume that a CAC precedes the association decision control.
A connection admission decision is taken by the CAC controller
before any mobile is considered as a candidate to connect to either
of the WLAN or UMTS networks. Thereafter, an association decision
only ensures global or individual optimal performance and it is not
proposed as an alternative to the CAC decision. However, the
association decision controller can still reject mobiles for optimal
performance of the network.

In our model, we introduce certain simplifying assumptions, as
compared to a real life scenario, in order to gain an analytical
insight into the dynamics of user-network association. Without these
assumptions it may be very hard to study these dynamics in a
WLAN-UMTS hybrid network.

\subsection{Related Work and Contributions}
\label{related-work}
Study of WLAN-UMTS hybrid networks is an emerging area of research
and not much related work is available. Authors in some related
papers (\cite{LFVT,JSD,CLCZ,MJ,NSI,HKAJ,OFHC}) have studied issues such as
vertical handover and coupling schemes, integrated architecture layout,
radio resource management (RRM) and mobility management. However, questions related to
load balancing or optimal user-network association have not been explored much.
Premkumar et al. in \cite{KPAK} propose a {\it near optimal}
solution for a hybrid network, within a
combinatorial optimization framework which is different from our approach.
To the best of our knowledge, ours is the first attempt to explicitly compute globally
optimal user-network association policies for a WLAN-UMTS hybrid network, under an
SMDP decision control formulation.
Moreover, this work is the first we know of to use stochastic
non-cooperative game theory to predict user behavior in
a decentralized decision making situation.
\junk{Moreover, we consider an alternate individual optimality
approach where we explicitly derive a system of linear equations to obtain the
expected service time of a mobile in a shared WLAN.}

\section{Model Framework}
\label{model-framework}

\junk{Consider only a single WLAN AP and a single UMTS NodeB present
in a geographical zone of arrival of mobile wireless devices. The AP
may be surrounded by other APs in its vicinity and the NodeB under
consideration is assumed to be surrounded by neighboring NodeBs as
in a standard cellular coverage network.}

A hybrid network may be composed of several 802.11 WLAN Access
Points (APs) and 3G-UMTS Base Stations (NodeBs) that are operated by
a single network operator. However, our focus is only on a single
pair of an AP and a NodeB that are located sufficiently close to
each other so that mobile users arriving in the combined coverage
area of this AP-NodeB pair, have a choice to connect to either of
the two networks. We call the combined coverage area network of a
single AP cell and a single NodeB micro-cell as a {\it hybrid cell}.
The cell coverage radius of a UMTS micro-cell is usually around $400
m$ to $1000 m$ whereas that of a WLAN cell varies from a few tens to
a few hundreds of meters. Therefore some mobiles arriving in the
hybrid cell may only be able to connect to the NodeB either because
they fall outside the transmission range of the AP or they are
equipped with only 3G technology electronics. While, other mobiles
that are equipped with only 802.11 technology can connect
exclusively to the WLAN AP. Apart from these two categories, mobiles
equipped with both 802.11 WLAN and 3G-UMTS technologies can connect
to any one of the two networks. The decision to connect to either of
the two networks can involve different cost or utility criteria. A
cost criteria could be the average service time of a mobile and an
example utility could comprise the throughput of a mobile. Moreover,
the connection or association decision involves two different
decision makers, the mobile user and the network operator. Leaving
the decision choice with the mobile user may result in less
efficient use of the network resources, but may be much more
scalable and easier to implement. We thus model the decision problem
in two different and alternate ways. Firstly, we consider the Global
Optimality dynamic control formulation in which the network operator
dictates the decision of mobile users to connect to one of the two
networks, so as to optimize a certain global cell utility. And
secondly, we consider the Individual Optimality dynamic control
formulation in which a mobile user takes a selfish decision to
connect to either of the two networks so that only its own cost is
optimized. We model the Global optimality problem with an SMDP (Semi
Markov Decision Process) control approach and the Individual
optimality problem under a non-cooperative dynamic game framework.
Before discussing further the two approaches, we first describe
below a general framework common to both. We also state some
simplifying assumptions and expressions for the downlink throughput
from previous work. Since the bulk of data transfer for a mobile
engaged in streaming or interactive data transmission is carried
over the downlink (AP to mobile or NodeB to mobile), we are
interested here in the TCP throughput of only downlink.

\junk{Note that in this paper we particularly study the two
approaches using these throughput expressions, however, results
obtained for both approaches are more generic and can be studied
with any other valid expressions for the throughputs.}
\subsection{Mobile Arrivals}
\label{mobile-arrivals}

We model the hybrid cell of an 802.11 WLAN AP and a 3G-UMTS NodeB as
an $M/G/2$ processing server system (Figures \ref{hybrid-net} \& \ref{hybrid-net-indiv}) with
each server having a separate finite pole capacity of $M_{AP}$ and
$M_{3G}$ mobiles, respectively. We will give further clarifications
on the pole capacity of each server later in Sections
\ref{thpt_exp_ap} and \ref{thpt_exp_nodeb}. As discussed
previously, mobiles are considered as candidates to connect to the
hybrid cell only after being admitted by a CAC, such as the one described
in \cite{whitepaper}. Some of the
admitted mobiles can connect only to the WLAN AP and some others
only to the 3G-UMTS NodeB. These two set of arriving mobiles are
each assumed to constitute two separate dedicated arrival streams
with Poisson rates $\lambda_{AP}$ and $\lambda_{3G}$, respectively.
The remaining set of mobiles which can connect to both networks
form a common arrival stream with Poisson rate $\lambda_{AP3G}$. The
mobiles of the two dedicated streams can either directly join their respective
AP or NodeB network without any connection decision
choice involved, or they can be rejected. For mobiles of common stream,
either a rejection or a connection routing decision has to be taken, as to
which of the two networks will the arriving mobiles join, while optimizing a certain cost or
utility. It is assumed that all arriving mobiles have a downlink data
service requirement which is exponentially distributed with
parameter $\zeta$. In other words, every arriving mobile seeks to
download a data file of average size $1/\zeta$ bits on the
downlink. Let $\theta_{AP}(m_c)$ denote the downlink throughput of
each mobile in the AP network when $m_c$ mobiles are connected to it
at any given instant. If $\eta_{DL}$ denotes the downlink {\it cell
load} of the NodeB cell, then assuming $N$ active mobiles to be
connected to the NodeB, $\eta \defin \frac{\eta_{DL}}{N}$ denotes
the average {\it load per user} in the cell. Let $\theta_{3G}(\eta)$ denote
the downlink throughput of each mobile in the NodeB network when its
average load per user is $\eta$. With the above notations,
the effective service rates of each network or server can be denoted
by $\mu_{AP}(m_c)=\zeta \times \theta_{AP}(m_c)$ and
$\mu_{3G}(\eta)=\zeta \times \theta_{3G}(\eta)$.

\subsection{Simplifying Assumptions}
\label{assumptions}

We assume a single QoS class of arriving mobiles so that each mobile
has an identical minimum downlink throughput requirement of
$\theta_{min}$, i.e., each arriving mobile must achieve a downlink
throughput of at least $\theta_{min}$ bps on either of the two
networks. It is further assumed that each mobile's or receiver's
advertised window $W^*$ is set to $1$ in the TCP protocol.
This is known to provide the best performance of TCP (see
\cite{ZPHSLM}, \cite{LP1} and references therein).

We further assume {\it saturated resource allocation} in the
downlink of AP and NodeB networks. Specifically, this assumption for
the AP network means the following. Assume that the AP is {\it
saturated} and has infinitely many packets backlogged in its
transmission buffer. In other words, there is always a packet in the
AP's transmission buffer waiting to be transmitted to each of the
connected mobiles. Now in a WLAN cell, resource allocation to an AP
on the downlink is carried out through the contention based DCF
(Distributed Coordination Function) protocol. If the AP is saturated
for a particular mobile's connection and $W^*$ is set to $1$, then
this particular mobile can benefit from higher number of
transmission opportunities ({\it TxOPs}) won by the AP for downlink
transmission to this mobile (hence higher downlink throughput), than
if the AP is not saturated or $W^*$ is not set to $1$. Thus with the
above assumptions, mobiles can be allocated downlink throughputs
greater than their QoS requirements of $\theta_{min}$ and cell
resources in terms of {\it TxOPs} on the downlink will be maximally
utilized.

For the NodeB network, the saturated resource allocation assumption
has the following elaboration. It is assumed that at any given
instant, the NodeB cell resources on downlink are fully utilized
resulting in a constant maximum cell load of $\eta^{max}_{DL}$. This
is analogous to the maximal utilization of {\it TxOPs} in the AP
network discussed in the previous paragraph. With this maximum cell load assumption even
if a mobile has a minimum throughput requirement of only
$\theta_{min}$ bps, it can actually be allocated a higher throughput
if additional unutilized cell resources are available, so that the
cell load is always at its maximum of $\eta^{max}_{DL}$. If say a new mobile
$j$ arrives and if it is possible to accommodate its connection
while maintaining the QoS requirements of the presently connected
mobiles (this will be decided by the CAC), then the NodeB will
initiate a {\it renegotiation} of QoS attributes (or bearer attributes)
procedure with all the presently connected mobiles. All presently
connected mobiles will then be allocated a lower throughput than the
one prior to the set-up of mobile $j$'s connection. However, this
new lower throughput will still be higher than each mobile's QoS
requirement. This kind of a renegotiation of QoS attributes is
indeed possible in UMTS and it is one of its special features (see Chapter 7
in \cite{WCDMA}). Also note a very key point here that the average load per user
$\eta$ as defined previously in Section \ref{mobile-arrivals}, decreases with increasing number of
mobiles connected to the NodeB. Though the total cell load is always
at its maximum of $\eta^{max}_{DL}$, contribution to this total load
from a single mobile (i.e., load per user, $\eta$) decreases as more
mobiles connect to the NodeB cell. We define $\Delta(\eta)$ as the average
change in $\eta$ caused by a new mobile that connects to the
NodeB cell. Therefore, when a new mobile connects, the load per user
drops from $\eta$ to $\eta - \Delta(\eta)$ and when a mobile
disconnects, the load per user increases from $\eta$ to $\eta +
\Delta(\eta)$.

In downlink, the inter-cell to intra-cell interference ratio denoted
by $i_j$ and the orthogonality factor denoted by $\alpha_j$ are
different for each mobile $j$ depending on its location in the NodeB
cell. Moreover, the throughput achieved by each mobile is
interference limited and depends on the signal to interference plus
noise ratio (SINR) received at that mobile. Thus, in the absence of
any power control, the throughput also depends on the location of
mobile in the NodeB cell. We assume a uniform SINR scenario where
closed-loop fast power control is applied in the NodeB cell, so that
each mobile receives approximately the same SINR. We therefore
assume that all mobiles in the NodeB cell are allocated equal
throughputs. This kind of a power control will allocate more power
to users far away from the NodeB that are subject to higher
path-loss, fading and neighboring cell interference. Users closer to
the NodeB will be allocated relatively less power since they are
susceptible to weaker signal attenuation. In fact, such a fair
throughput allocation can also be achieved by adopting a fair and
power-efficient channel dependent scheduling scheme as described in
\cite{ZHZ03}. Now since all mobiles are allocated equal throughputs,
it can be said that mobiles arrive at an {\em average} location in the
NodeB cell (see Section 8.2.2.2 in \cite{WCDMA}). Therefore all
mobiles are assumed to have an identical average inter-cell to
intra-cell interference ratio $\bar{i}$ and an identical average
orthogonality factor $\bar{\alpha}$.

The assumption on saturated resource allocation is a standard
assumption, usually adopted to simplify modeling of complex network
frameworks like those of WLAN and UMTS (see for e.g.,
\cite{WCDMA,KAMG04}). Mobiles in NodeB cell are assumed to be
allocated equal throughputs in order to have a comparable scenario
to that of an AP cell, in which mobiles are also known to achieve
fair and equal throughput allocation (see Section
\ref{thpt_exp_ap}). Moreover such fair throughput allocation is
known to result in a better delay performance for typical file
transfers in UMTS (see \cite{TB1}). Furthermore, the assumption of
mobiles arriving at an average location in the NodeB cell, is
essential in order to simplify our models in Sections
\ref{semi-mdp-control} and \ref{Individual-Optimality}. For
instance, in the global optimality model, without this assumption
the hybrid network system state will have to include the location of
each mobile. This will result in a higher dimensional SMDP problem
which is analytically intractable.

\begin{figure}[!t]
\begin{minipage}{5in}
\centering
\includegraphics[width=2.8in]{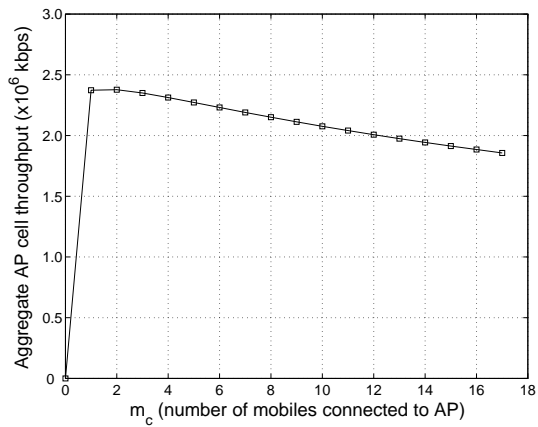}
\caption{Total throughput of all mobiles in an AP cell}
\label{theta_ap}
\end{minipage}
\end{figure}

\subsection{Downlink Throughput in 802.11 WLAN AP}
\label{thpt_exp_ap}

We reuse the downlink TCP throughput formula for a mobile in a WLAN
from \cite{MKA06}. For completeness, here we briefly mention the
network model that has been extensively studied in \cite{MKA06} and
then simply restate the throughput expression without going into
much details. Each mobile connected to the AP uses the Distributed
Coordination Function (DCF) protocol with an RTS/CTS frame exchange
before any data-ack frame exchange and each mobile has an equal
probability of the channel being allocated to it. With the
assumption of $W^*$ being set to $1$ (Section \ref{assumptions}) any
mobile will always have a TCP ack waiting to be sent back to the AP
with probability $1/2$, which is also the probability that it
contends for the channel. This is however true only for those
versions of TCP that do not use delayed acks. If the AP is always
saturated or backlogged, the average number of backlogged mobiles
contending for the channel is given by $m_b=1+\frac{m_c}{2}$. Based
on this assumption and since for any connection an ack is sent by
the mobile for every TCP packet received, the downlink TCP
throughput of a single mobile is given by Section 3.2 in
\cite{MKA06} as,
\begin{equation}
\label{WLAN_thpt}
\theta_{AP}(m_c) = \frac{L_{TCP}}{m_c (T_{TCPdata}+T_{TCPack}+2T_{tbo}+2T_w)},
\end{equation}
where $L_{TCP}$ is the size of TCP packets and $T_{TCPdata}$ and
$T_{TCPack}$ are the raw transmission times of a TCP data and a TCP
ack packet, respectively. $T_{tbo}$ and $T_w$ denote the mean total
time spent in {\it back-off} and the average total time wasted in
collisions for any successful packet transmission and are computed
assuming $m_b$ backlogged mobiles. The explicit expressions for
$T_{TCPdata}$, $T_{TCPack}$, $T_{tbo}$ and $T_w$ can be referred to
in \cite{MKA06}. However, we mention here that they depend on
certain quantities whose numerical values have been provided in
Section \ref{semi-mdp-control}. Note that all mobiles connected to
the AP achieve equal downlink TCP throughputs in a fair manner,
given by Equation \ref{WLAN_thpt}. Figure \ref{theta_ap} shows a
plot of total cell throughput in an AP cell. Since the total
throughput monotonically decreases with increasing number of
mobiles, the pole capacity of an AP cell $M_{AP}$ is limited by the
QoS requirement $\theta_{min}$ bps of each mobile.

\begin{figure}[!t]
\begin{minipage}{5in}
\centering
\includegraphics[width=2.8in]{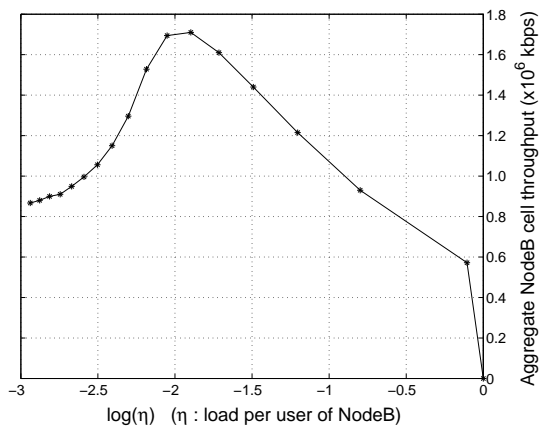}
\caption{Total throughput of all mobiles in NodeB cell}
\label{theta_nodeb}
\end{minipage}
\end{figure}

\subsection{Downlink Throughput in 3G-UMTS NodeB}
\label{thpt_exp_nodeb} We consider a standard model for data
transmission on downlink in a 3G-UMTS NodeB cell. Let $W$ be the
WCDMA modulation bandwidth and if $SINR$ denotes the signal to
interference plus noise ratio received at a mobile then its energy
per bit to noise density ratio is given by,
\begin{equation}
\label{eq:EbNo} \frac{E_b}{N_o} = \frac{W}{\theta_{3G}} \times SINR.
\end{equation}
Now, under the assumptions of identical throughput allocation to
each mobile arriving at an average location and application of power
control so that each mobile receives the same SINR (Section
\ref{assumptions}), we deduce from Eq. \ref{eq:EbNo} that each
mobile requires the same $E_b/N_o$ ratio in order to be able to
successfully decode NodeB's transmission. From Chapter 8 in
\cite{WCDMA} we can thus say that the downlink TCP throughput
$\theta_{3G}$ of any mobile, in a NodeB cell with saturated resource
allocation, as a function of load per user $\eta$ is given by,
\begin{equation}
\label{UMTS_thpt} \theta_{3G}(\eta)= \frac{\eta
W}{(E_b/N_o)(1-\bar{\alpha}+\bar{i})},
\end{equation}
where $\bar{\alpha}$ and $\bar{i}$ have been defined before in
Section \ref{assumptions}.
Figure \ref{theta_nodeb} shows a plot of total cell throughput of
all mobiles against $log(\eta)$ in a UMTS NodeB cell. The load per
user $\eta$ has been stretched to a logarithmic scale for better
presentation. Also note that throughput values have been plotted in
the second quadrant. As we go away from origin on the horizontal
axis, $log(\eta)$ (and $\eta$) decreases or equivalently number of
connected mobiles increase. The equivalence between $\eta$ and
$log(\eta)$ scales and number of mobiles $N(\eta)$ can be referred to
in Table \ref{table_quant}.

\begin{table}
\begin{scriptsize}
\[ \begin{array}{||c|c|c||c||c||c||}
\hline \eta & log(\eta) & N(\eta) & SINR & \theta_{3G}& \frac{E_b}{N_o}\\
& & & (dB) & (kbps) & (dB)\\
\hline 0.9    & -0.10536  & 1  & 0.8423   & 572 & 9.0612 \\
\hline 0.45   & -0.79851  & 2  & -2.1804  & 465 & 6.9503 \\
\hline 0.3    & -1.204    & 3  & -3.7341  & 405 & 5.7894 \\
\hline 0.225  & -1.4917   & 4  & -5.1034  & 360 & 5.0515 \\
\hline 0.18   & -1.7148   & 5  & -6.0327  & 322 & 4.5669 \\
\hline 0.15   & -1.8971   & 6  & -6.5093  & 285 & 4.3052 \\
\hline 0.1286 & -2.0513   & 7  & -7.2075  & 242 & 4.3460 \\
\hline 0.1125 & -2.1848   & 8  & -8.8312  & 191 & 4.7939 \\
\hline 0.1    & -2.3026   & 9  & -8.9641  & 144 & 5.5091 \\
\hline 0.09   & -2.4079   & 10 & -9.1832  & 115 & 6.0281 \\
\hline 0.0818 & -2.5033   & 11 & -9.9324  & 96  & 6.3985 \\
\hline 0.0750 & -2.5903   & 12 & -10.1847    & 83  & 6.6525 \\
\hline 0.0692 & -2.6703   & 13 & -10.7294 & 73  & 6.8625 \\
\hline 0.0643 & -2.7444   & 14 & -10.9023 & 65  & 7.0447 \\
\hline 0.06   & -2.8134   & 15 & -10.9983 & 60  & 7.0927 \\
\hline 0.0563 & -2.8779   & 16 & -11.1832 & 55  & 7.1903 \\
\hline 0.0529 & -2.9386   & 17 & -11.3802 & 51  & 7.2549 \\
\hline 0.05   & -2.9957   & 18 & -11.9231 & 47  & 7.3614 \\
\hline
\end{array}\]
\caption{} \label{table_quant}
\end{scriptsize}
\end{table}

It is to be noted here that the required $E_b/N_o$ ratio by each
mobile is a function of its throughput. Also, if the NodeB cell is
fully loaded with $\eta_{DL}=\eta^{max}_{DL}$ and if each mobile operates at its
minimum throughput requirement of $\theta_{min}$ then we can easily
compute the pole capacity $M_{3G}$ of the cell as,
\begin{equation}
\label{UMTS_pole_cap} M_{3G} = \frac{\eta^{max}_{DL}
W}{\theta_{min}(E_b/N_o)(1-\bar{\alpha}+\bar{i})}.
\end{equation}
For $\eta^{max}_{DL}=0.9$ and a typical NodeB cell scenario that
employs the closed-loop fast power control mechanism mentioned
previously in Section \ref{assumptions}, Table \ref{table_quant}
shows the SINR (fourth column) received at each mobile as a function
of the avg. load per user (first column). Note that we consider a
maximum cell load of $0.9$ and not $1$ in order to avoid instability
conditions in the cell. These values of SINR have been obtained from
radio layer simulations of a NodeB cell. The values shown here have
been slightly modified since the original values are part of a
confidential internal document at France Telecom R\&D. The fifth
column shows the downlink throughput with a block error rate (BLER)
of $10^{-2}$ that can be achieved by each mobile as a function of
the SINR observed at that mobile. And the sixth column in the table
lists the corresponding values of $E_b/N_o$ ratio (obtained from
Equation \ref{eq:EbNo}), that are required at each mobile to
successfully decode NodeB's transmission.

\begin{figure}
\begin{minipage}{5in}
\centering
\psfrag{ETA}{$\eta$}
\psfrag{MUC}{$m_c$}
\psfrag{INN}{$\in$}
\psfrag{TOO}{$\to$}
\includegraphics[width=3.4in]{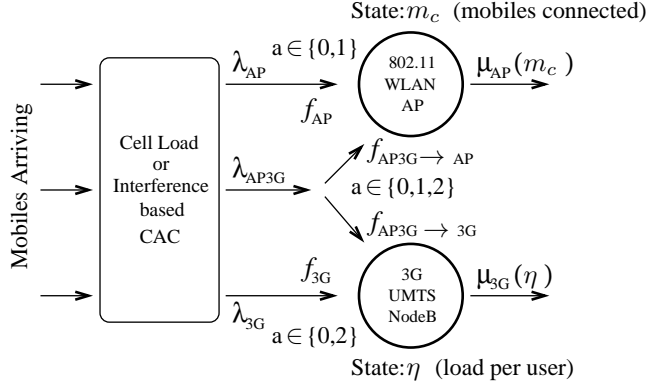}
\caption{Hybrid cell scenario under Global optimality}
\label{hybrid-net}
\end{minipage}
\end{figure}

\section{Global Optimality: SMDP control formulation}
\label{semi-mdp-control} In the Global Optimality approach, it is
the network operator that takes the optimal decision for each mobile
as to which of the two AP or NodeB networks the mobile will connect
to, after it has been admitted into the hybrid cell by the CAC
controller (Figure \ref{hybrid-net}). Since decisions have to be
made at each arrival, this gives an SMDP structure to the decision
problem and we state the equivalent SMDP problem as follows:

\begin{itemize}
\item {\it States:} The state of a hybrid cell system is denoted by the tuple $(m_c,\eta)$
where $m_c$ $(0\leq m_c \leq M_{AP})$ denotes the number of mobiles
connected to the AP and $\eta$ $(0.05 \le \eta \le 0.9)$ is the load
per user of the NodeB cell.
\item {\it Events:} We consider two distinguishable events: (i) arrival
of a new mobile after it has been admitted by CAC and (ii) departure
of a mobile after service completion.
\item {\it Decisions:} For mobiles arriving in the common stream a decision
action $a \in \{0,1,2\}$ has to be taken. $a=0$ represents rejecting
the mobile, $a=1$ represents routing the mobile connection to AP
network and $a=2$ represents routing the mobile connection to NodeB
network.
\item {\it Rewards:} Whenever a new incoming mobile is either
rejected or routed to one of the two networks, it generates a
certain state dependent reward. Generally, the aggregate throughput of an AP or
NodeB cell drops when an additional new mobile connects to it.
However the network operator gains some financial revenue from the
mobile user at the same time. There is thus a trade-off between
revenue gain and the aggregate network throughput which motivates us
to formulate the reward as follows. The reward consists of the sum
of a fixed financial revenue price component and $\beta$ times an
aggregate network throughput component, where $\beta$ is an
appropriate proportionality constant. When a mobile of the dedicated
arrival streams is routed to the corresponding AP or NodeB, it
generates a financial revenue of $f_{AP}$ and $f_{3G}$,
respectively. A mobile of the common stream generates a financial
revenue of $f_{AP3G\to AP}$ on being routed to the AP and
$f_{AP3G\to 3G}$ on being routed to the NodeB. Any mobile
that is rejected does not generate any financial revenue. The
throughput component of the reward is represented by the aggregate
network throughput of the corresponding AP or NodeB network to which a newly
arrived mobile connects, taking into account the change in the state
of the system caused by this new mobile's connection. Whereas, if a
newly arrived mobile in a dedicated stream is rejected then the
throughput component represents the aggregate network throughput of
the corresponding AP or NodeB network, taking into account the unchanged state
of the system. For a rejected mobile belonging to the common stream,
it is the maximum of the aggregate network throughputs of the two
networks that is considered.
\item {\it Criterion:} The optimality criterion is to maximize the total expected
discounted reward over an infinite horizon and obtain a
{\it deterministic} and {\it stationary} optimal policy.
\end{itemize}

\junk{ When a new mobile is routed to network $1$ with $i$ mobiles
already connected to it, each mobiles' throughput changes from
$\theta_{AP}(i)$ to $\theta_{AP}(i+1)$. Thus a cost of
$\theta_{AP}(i)-\theta_{AP}(i+1)$ is incurred on taking action $a=1$
and a cost of $\theta_{3G}(j)-\theta_{3G}(j+1)$ is incurred on
taking action $a=2$. Equivalently, we can say that actions $a=1$ and
$a=2$ yield a reward of $-(\theta_{AP}(i)-\theta_{AP}(i+1))$ and
$-(\theta_{3G}(j)-\theta_{3G}(j+1))$, respectively. }

Note that in the SMDP problem statement above, state transition
probabilities have not been mentioned because depending on the
action taken, the system moves into a unique new state
deterministically, i.e., w.p. 1. For instance when action $a=1$ is
taken, the state evolves from $(m_c,\eta)$ to $(m_c+1,\eta)$ or when
action $a=2$ is taken, the state evolves from $(m_c,\eta)$ to
$(m_c,\eta-\Delta(\eta))$. Applying the well-known {\it
uniformization} technique from \cite{lippman}, we can say that
events (i.e., arrival or departure of mobiles) occur at the jump
times of the combined Poisson process of all types of events with
rate $\Lambda:=\lambda_{AP} + \lambda_{3G} + \lambda_{AP3G} +
{\check \mu}_{AP} + {\check \mu}_{3G}$, where ${\check \mu}_{AP} :=
\max_{m_c}{\mu_{AP}(m_c)}$ and ${\check \mu}_{3G} :=
\max_{\eta}{\mu_{3G}(\eta)}$. The departure of a mobile is either a
real departure, or an {\it artificial} departure, when from a single
mobile's point of view the corresponding server slows down due to
large number of mobiles in the network. Then, any event occurring,
corresponds to an arrival on the dedicated streams with probability
$\lambda_{AP} / \Lambda$ and $\lambda_{3G} / \Lambda$, an arrival on
the common stream with probability $\lambda_{AP3G} / \Lambda$ and a
real departure with probability $\mu_{AP}(m_c) / \Lambda$ or
$\mu_{3G}(\eta) / \Lambda$. As a result, the time {\it periods}
between consecutive events are i.i.d. distributed and we can
consider an $n-$stage SMDP decision problem. Let $V_n(m_c,\eta)$
denote the maximum expected $n-$stage discounted reward for the
hybrid cell, when the system is in state $(m_c,\eta)$. The
stationary optimal policy that achieves the maximum total expected
discounted reward over an infinite horizon can then be obtained as a
solution of the $n-$stage problem as $n\to \infty$. The discount
factor is denoted by $\gamma$ ($0<\gamma<1$) and determines the
relative worth of the present reward v/s the future rewards. State
$(m_c,\eta)$ of the system is observed right after the occurrence of
an event, for example, right after a newly arriving mobile in the
common stream has been routed to one of the networks, or right after
the departure of a mobile. Let $U_n(m_c,\eta;a)$ denote the maximum
expected $n-$stage discounted reward for the hybrid cell when the
system is in state $(m_c,\eta)$, given that an arrival event has
occurred and given that action '$a$' will be taken for this newly
arrived mobile. We can then write down the following recursive
Dynamic Programming (DP) equation to solve our SMDP decision
problem. $\forall n\geq 0$ and $0\leq m_c \leq M_{AP}$, $0.05 \leq
\eta \leq 0.9$,

\begin{equation}
\begin{split}
V_{n+1}(m_c,\eta) &  =  \frac{\lambda_{AP}}{\Lambda}
\max_{a\in \{0,1\}} \left\{ R_{AP}(m_c,\eta;a) + \gamma U_n(m_c,\eta;a) \right\} \\
& \quad + \frac{\lambda_{3G}}{\Lambda} \max_{a\in \{0,2\}}
\left\{ R_{3G}(m_c,\eta;a) + \gamma U_n(m_c,\eta;a) \right\} \\
& \quad + \frac{\lambda_{AP3G}}{\Lambda} \max_{a\in
\{0,1,2\}}
\left\{ R_{AP3G}(m_c,\eta;a) + \gamma U_n(m_c,\eta;a) \right\} \\
& \quad + \frac{\mu_{AP}(m_c)}{\Lambda} \gamma V_n( (m_c-1)\vee 0,\eta) \\
& \quad + \frac{\mu_{3G}(\eta)}{\Lambda} \gamma V_n(m_c,(\eta+\Delta(\eta))\wedge0.9 ) \\
& \quad + \frac{\Lambda -(\lambda_{AP}+\lambda_{3G}+\lambda_{AP3G}+\mu_{AP}(m_c)+\mu_{3G}(\eta))}{\Lambda} \gamma V_n(m_c,\eta), \\
\end{split}
\end{equation}
where,
\begin{equation}
\label{eq:reward1}
\begin{split}
R_{AP}(m_c,\eta;a) & = \left\{ \begin{array}{cc} \beta \; m_c \; \theta_{AP}(m_c) & \hspace{-1.85cm} \;\;: a=0 \\
f_{AP} + \beta \; (m_c+1) \; \theta_{AP}(m_c+1) & \;\; : a=1, m_c< M_{AP} \\
\beta \; m_c \; \theta_{AP}(m_c) & \;\; : a=1, m_c=M_{AP} \end{array}\right.
\end{split}
\end{equation}
\begin{equation}
\label{eq:reward2}
\begin{split}
\hspace{0.25cm}
R_{3G}(m_c,\eta;a) & = \left\{ \begin{array}{cc} \beta \; N(\eta) \; \theta_{3G}(\eta) & \hspace{-2.15cm} \;\; : a=0 \\
f_{3G} + \beta \; N(\eta-\Delta(\eta)) \; \theta_{3G}(\eta-\Delta(\eta)) & \;\; : a=2, N(\eta)<M_{3G} \\
\beta \; N(\eta) \; \theta_{3G}(\eta) & \;\; : a=2, N(\eta)=M_{3G} \end{array}\right.
\end{split}
\end{equation}
\begin{equation}
\label{eq:reward1}
\begin{split}
R_{AP3G}(m_c,\eta;a) & = \left\{ \begin{array}{cc} \max\{\beta \; m_c \; \theta_{AP}(m_c), \beta \; N(\eta) \; \theta_{3G}(\eta)\} & \hspace{-1.85cm} \;\; : a=0 \\
f_{AP3G\to AP} + \beta \; (m_c+1) \; \theta_{AP}(m_c+1) & \;\; : a=1, m_c<M_{AP} \\
\beta \; m_c \; \theta_{AP}(m_c) & \;\; : a=1, m_c=M_{AP} \\
f_{AP3G\to 3G} + \beta \; N(\eta-\Delta(\eta)) \; \theta_{3G}(\eta-\Delta(\eta)) & \hspace{0.25cm} \;\; : a=2, N(\eta)<M_{3G} \\
\beta \; N(\eta) \; \theta_{3G}(\eta) & \hspace{0.25cm} \;\;: a=2, N(\eta)=M_{3G} \end{array}\right.
\end{split}
\end{equation}
\noindent
and $U_n(m_c,\eta;0) := V_n(m_c,\eta)$, $U_n(m_c,\eta;1) :=
V_n((m_c+1)\wedge M_{AP},\eta)$, $U_n(m_c,\eta;2) :=
V_n(m_c,(\eta-\Delta(\eta))\vee 0.05)$ for $\theta_{min}=46$ kbps
and $N(\eta)$ can be obtained
from Table \ref{table_quant}. We solve the above DP equation with
Value Iteration method using the following numerical values for
various entities: $L_{TCP} = 8000$ bits (size of TCP packets),
$L_{MAC} = 272$ bits, $L_{IPH} = 320$ bits (size of MAC and TCP/IP
headers), $L_{ACK} = 112$ bits (size of MAC layer ACK), $L_{RTS} =
180$ bits, $L_{CTS} = 112$ bits (size of RTS and CTS frames),
$R_{data} = 11$ Mbits/s, $R_{control} = 2$ Mbits/s (802.11 data
transmission and control rates), $CW_{min} = 32$ (minimum 802.11
contention window), $T_P = 144\mu s$, $T_{PHY} = 48\mu s$ (times to
transmit the PLCP preamble and PHY layer header), $T_{DIFS} = 50\mu
s$, $T_{SIFS} = 10\mu s$ (distributed inter-frame spacing time and
short inter-frame spacing time), $T_{slot} = 20\mu s$ (slot size
time), $K = 7$ ({\it retry limit} in 802.11 standard), $b_0 = 16$
(initial mean back-off), $p = 2$ (exponential back-off multiplier),
$\gamma=0.8$, $\lambda_{AP}=0.03$, $\lambda_{3G}=0.03$,
$\lambda_{AP3G}=0.01$, $\zeta=10^{-6}$, $\beta=10^{-6}$, $M_{AP}=18$
and $M_{3G}=18$ for $\theta_{min}=46$ kbps, $\bar{\alpha}=0.9$ for ITU
Pedestrian A channel, $\bar{i}=0.7$, $W=3.84$ Mcps and other values
as illustrated in Table \ref{table_quant}.

\begin{figure}
\begin{minipage}{4.62in}
\centering
\includegraphics[width=2.5in]{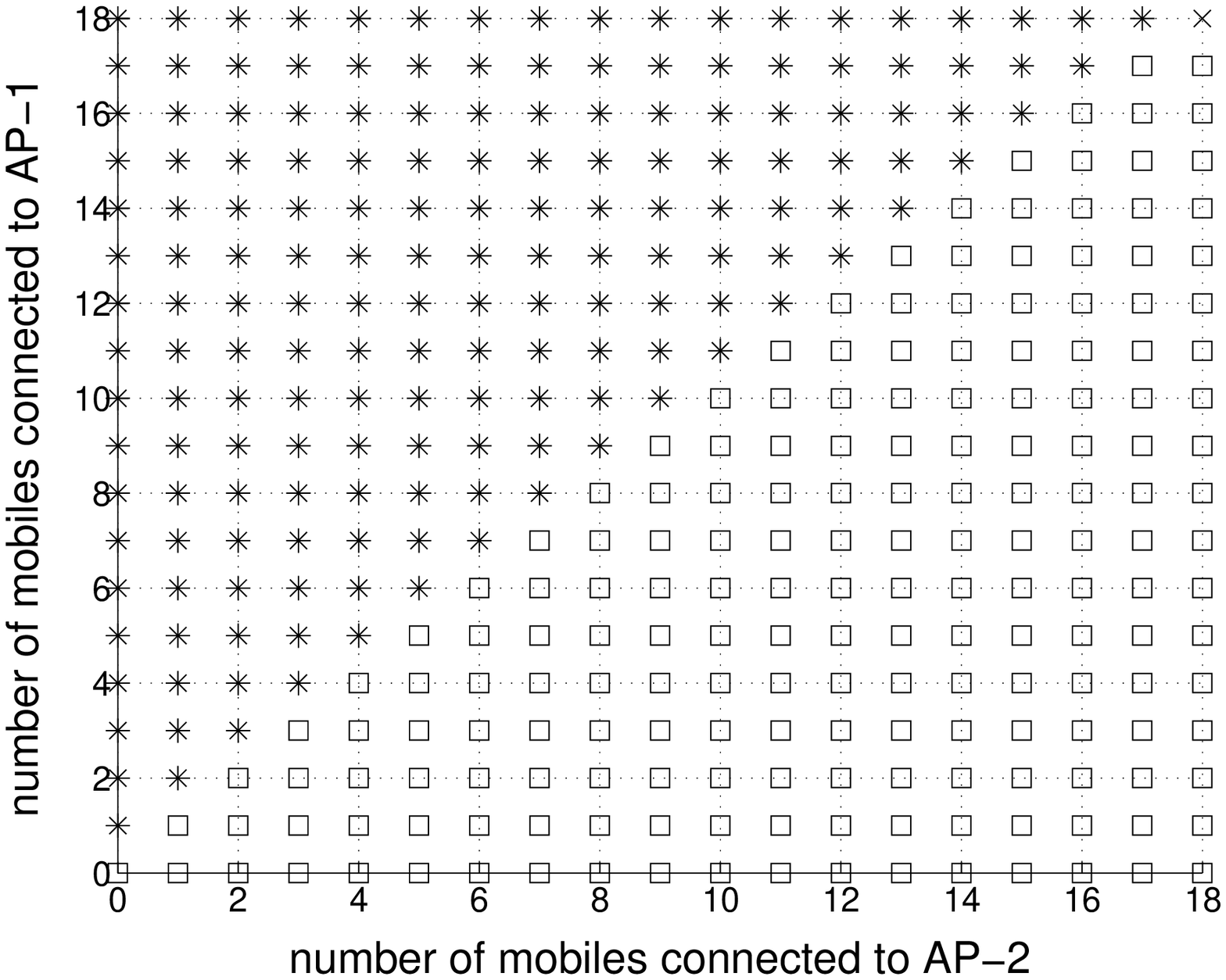}
\caption{Optimal policy for common flow in AP-AP setup.
{\it First} network: AP1, {\it Second} Network: AP2.}
\label{mc_mc}
\end{minipage}
\\
\begin{minipage}{5in}
\centering
\includegraphics[width=2.5in]{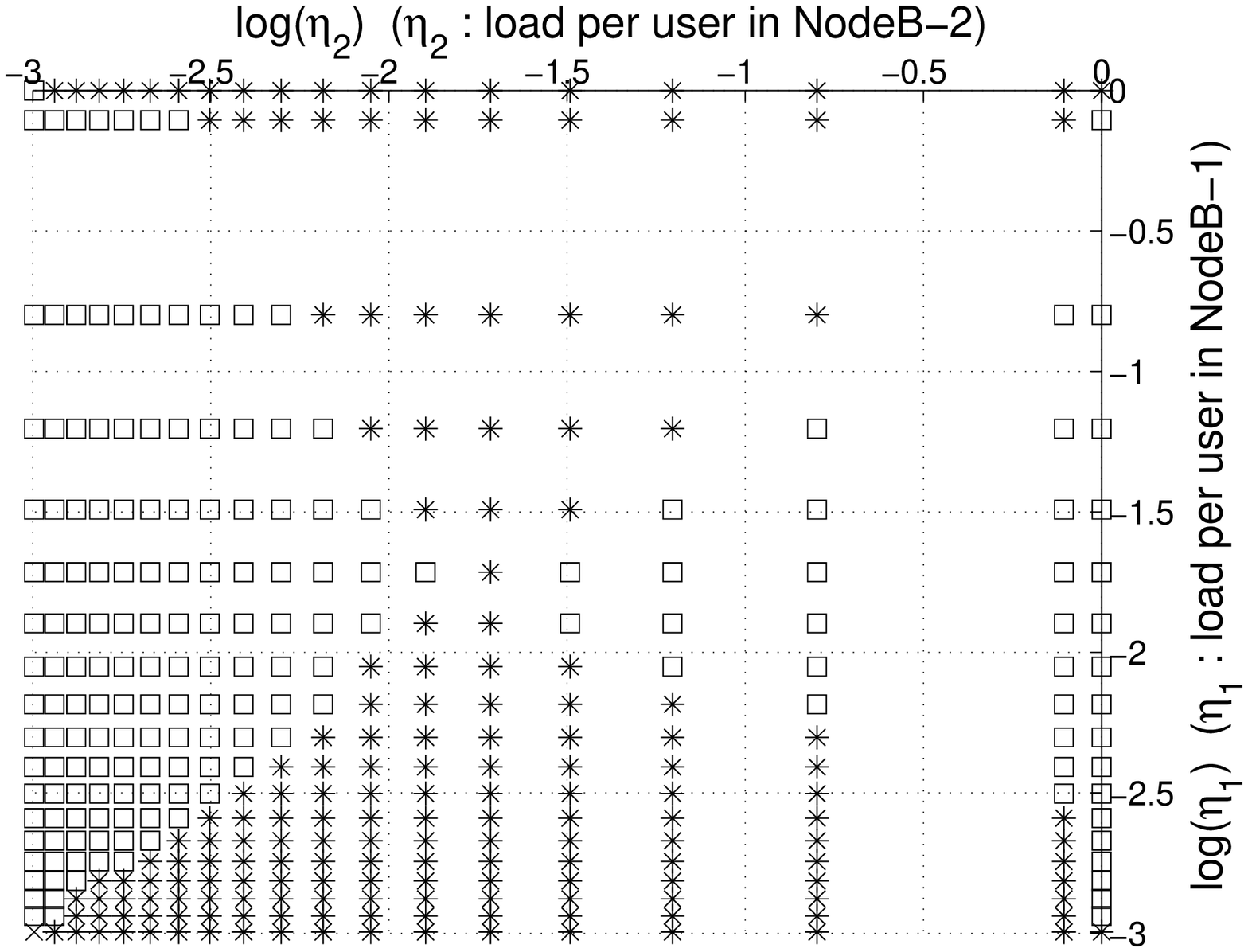}
\caption{Optimal policy for common flow in NodeB-NodeB setup.
{\it First} network: NodeB1, {\it Second} Network: NodeB2.}
\label{eta_eta}
\end{minipage}
\\
\begin{minipage}{5in}
\centering
\includegraphics[width=2.5in]{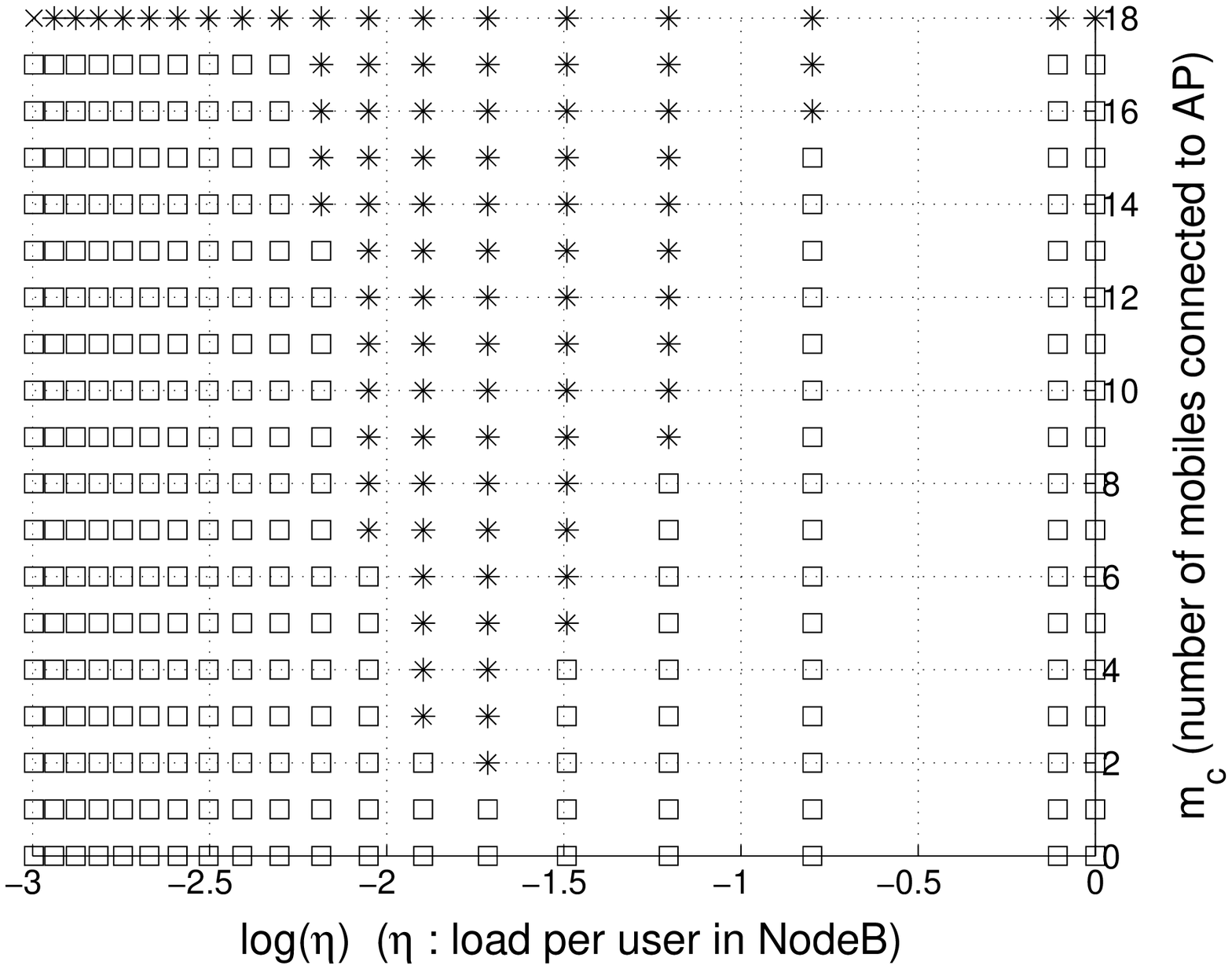}
\caption{Optimal policy for common flow in AP-NodeB hybrid cell.
{\it First} network: AP, {\it Second} Network: NodeB.}
\label{eta_mc}
\end{minipage}
\end{figure}

\begin{figure}
\begin{minipage}{5in}
\centering
\includegraphics[width=2.5in]{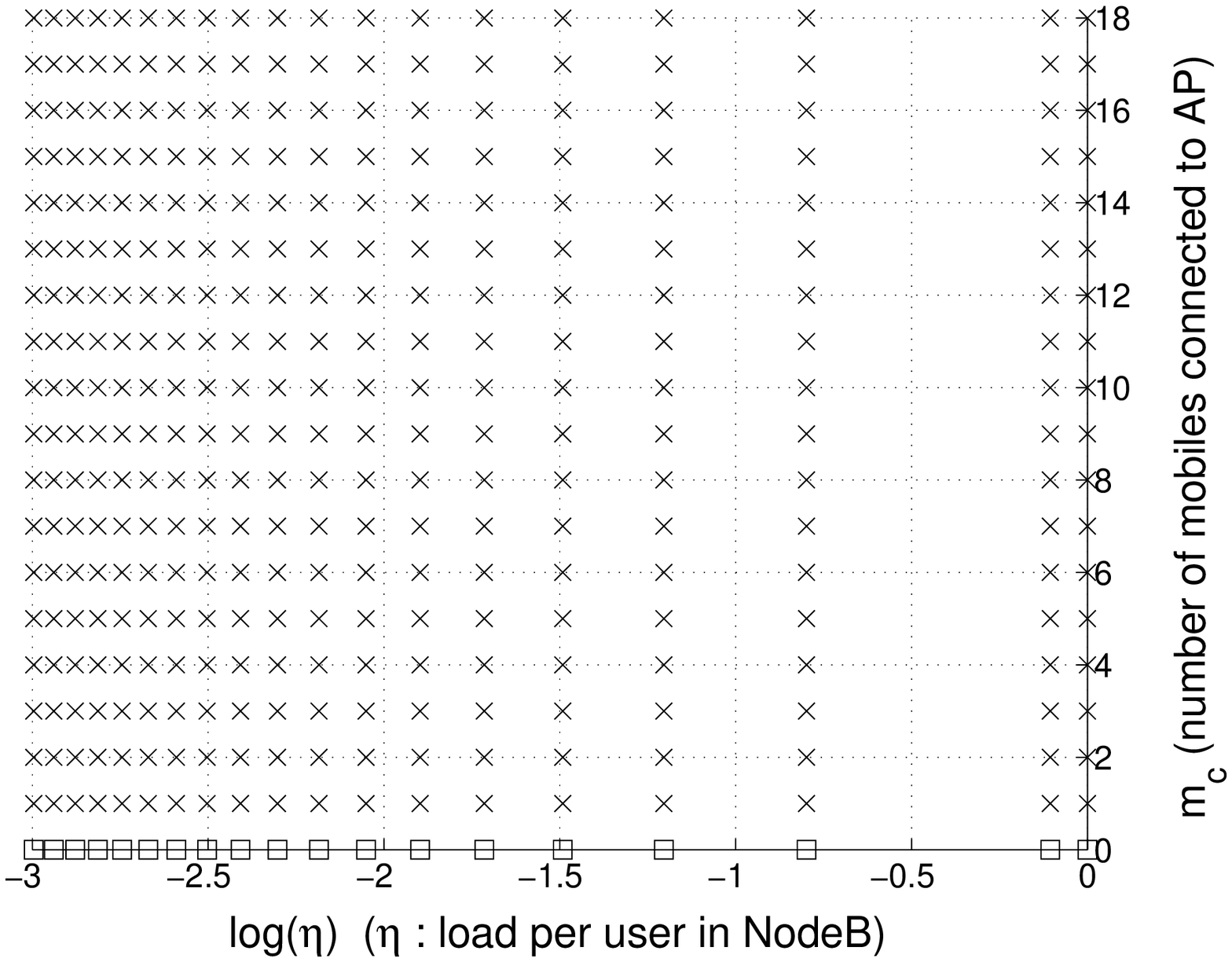}
\caption{Optimal policy for AP dedicated flow in AP-NodeB hybrid cell}
\vspace{0.2cm}
\label{eta_mc_1}
\end{minipage}
\begin{minipage}{5in}
\centering
\includegraphics[width=2.5in]{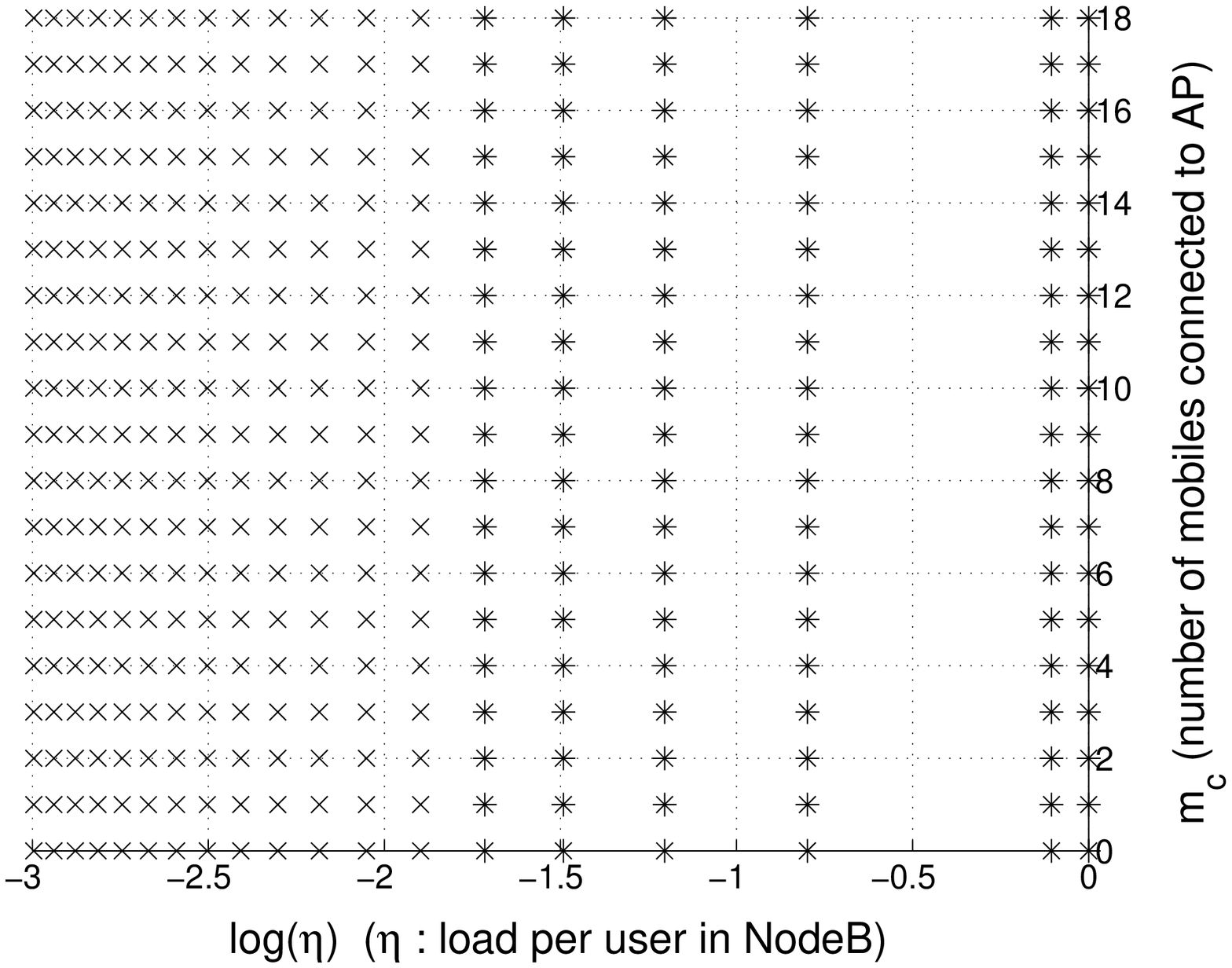}
\caption{Optimal policy for NodeB dedicated flow in AP-NodeB hybrid cell}
\vspace{0.2cm}
\label{eta_mc_2}
\end{minipage}
\end{figure}

The DP equation has been solved for three different kinds of
network setups. We first study the simple {\it homogenous} network case
where both networks are AP and hence an incoming mobile belonging to
the common stream must be offered a connection choice between two
identical AP networks. Next, we study an analogous case where both
networks are NodeB terminals. We study these two
cases in order to gain some insight into connection routing
dynamics in simple homogenous network setups before studying the
third more complex, hybrid AP-NodeB scenario. Figures \ref{mc_mc}-\ref{eta_mc_2}
show the optimal connection routing policy for the three network
setups. Note that the plot in Figure \ref{eta_eta} is in the $3^{rd}$
quadrant and plots in Figures \ref{eta_mc}-\ref{eta_mc_2} are in
the $2^{nd}$ quadrant. In all these figures a square box symbol ($\Box$) denotes
routing a mobile's connection to the {\it first} network, a star symbol
({$\ast$}) denotes routing to the {\it second} network and a cross
symbol ($\times$) denotes rejecting a mobile all together.

In Figure \ref{mc_mc}, optimal policy for the common stream
in an AP-AP homogenous network setup is shown
with $f_{AP1AP2\to AP1}=f_{AP1AP2\to AP2}=5$ (with some abuse of notation).
The optimal policy routes mobiles of common stream to the network which
has lesser number of mobiles than the other one. We refer to this behavior
as {\it mobile-balancing} network phenomenon. This happens because
the total throughput of an AP network decreases with increasing
number of mobiles (Figure \ref{theta_ap}). Therefore,
an AP network with higher number of mobiles offers lesser
reward in terms of network throughput and a mobile generates greater
incentive by joining the network with fewer mobiles.
Also note that the optimal routing policy in this case
is {\it symmetric} and of {\it threshold type} with the threshold switching
curve being the coordinate line $y=x$.

Figure \ref{eta_eta} shows the optimal routing policy for the common
stream in a NodeB-NodeB homogenous network setup. With equal financial
incentives for the mobiles, i.e., $f_{3G13G2\to
3G1}=f_{3G13G2\to 3G2}=5$ (with some abuse of notation), we observe
a very interesting switching curve structure. The state space in
Figure \ref{eta_eta} is divided into an {\it L-shaped} region (at bottom-left)
and a {\it quadrilateral shaped} region (at top-right) under the optimal policy. Each
region separately, is {\it symmetric} around the coordinate diagonal line $y=x$.
With some abuse of notation, consider the state $(\eta_1,\eta_2)=(-0.79851,-1.4917)$
(not the coordinate point) of the homogenous network on
logarithmic scale in the upper triangle of the quadrilateral region.
From Table \ref{table_quant} this corresponds to the network state
when load per user in the first NodeB network is $0.45$ which
is more than the load per user of $0.225$ in the second NodeB
network. Equivalently, there are less mobiles connected to the first
network as compared to the second network. Ideally, one would expect
new mobiles to be routed to the first network rather than the second
network. However, according to Figure \ref{eta_eta}, in this state the
optimal policy is to route to the second network even though the
number of mobiles connected to it is more than those in the first.
We refer to this behavior as {\it mobile-greedy} network phenomenon
and explain the intuition behind it in the following
paragraph. The routing policies on boundary coordinate lines are clearly
comprehensible. On $y=-2.9957$ line when
the first network is full (i.e., with least possible
load per user), incoming mobiles are routed to second network
(if possible) and vice-versa for the line $x=-2.9957$. When both
networks are full, incoming mobiles are rejected which is indicated
by the cross at coordinate point $(x,y)=(-2.9957,-2.9957)$.

The reason behind the mobile-greedy phenomenon in Figure
\ref{eta_eta} can be attributed to the fact that in a NodeB network,
the total throughput increases with decreasing avg. load per user up to a
particular threshold (say $\eta_{thres}$) and then decreases
thereafter (see Figure \ref{theta_nodeb}). Therefore, routing new
mobiles to a network with lesser (but greater than $\eta_{thres}$)
load per user results in a higher reward in terms of total network
throughput, than routing new mobiles to the other network with
greater load per user. However, the mobile-greedy phenomenon
is only limited to the quadrilateral shaped region. In the L-shaped
region, the throughput of a NodeB network decreases with decreasing
load per user, contrary to the quadrilateral region where the
throughput increases with decreasing load per user. Hence, in the
L-shaped region higher reward is obtained by routing to the network
having higher load per user (lesser number of mobiles) than by
routing to the network with lesser load per user (greater number of
mobiles). In this sense the L-shaped region shows similar characteristics
to mobile-balancing phenomenon observed in AP-AP network
setup (Figure \ref{mc_mc}).

\junk{ Hence it seems beneficial, in terms of our reward function
defined in Equations \ref{eq:reward1}-\ref{eq:reward3}, to accept
dedicated flow mobiles {\it only} until when there are $m$ mobiles
already connected to each of the respective networks and reject them
when there are more than $m$ mobiles connected.
-----------------
Figures \ref{w-w-9.00001-9}-\ref{w-w-9.001-9} show the optimal
routing policy again for hybrid flow 3 but with varying values of
$f_{31}$ as indicated below the figures. It can be seen that the
equilibrium between the L-shaped and the square shaped regions is
easily disturbed by only slightly changing the value of $f_{31}$. In
fact with increasing value of $f_{31}$, the optimal policy favors
routing of incoming mobiles to network 1 over network 2 due to
higher financial component of the reward obtained in network 1 than
that obtained in network 2.}

We finally discuss the hybrid AP-NodeB network setup. Here we consider
financial revenue gains of $f_{AP3G\to AP}=5$ and $f_{AP3G\to 3G}=5.65$,
motivated by the fact that a network operator can charge more for a
UMTS connection since it offers a larger coverage area and moreover
UMTS equipment is more expensive to install and maintain than WLAN
equipment. In Figure \ref{eta_mc}, we observe that
the state space is divided into two regions by the optimal
policy switching curve which is {\it neither convex nor concave}.
Moreover, in some regions of state space the mobile-balancing network
phenomenon is observed, where as in some other regions the mobile-greedy
network phenomenon is observed.
In some sense, this can be attributed to the symmetric threshold type switching
curve and the symmetric L-shaped and quadrilateral shaped regions in
the corresponding AP-AP and NodeB-NodeB homogenous network setups, respectively.
Figures \ref{eta_mc_1} and \ref{eta_mc_2} show the optimal policies
for dedicated streams in an AP-NodeB hybrid cell with
$f_{AP}=f_{3G}=0$. The optimal policy accepts new mobiles in the AP
network only when there are none already connected. This happens
because the network throughput of an AP is zero when there are no
mobiles connected and a non-zero reward is obtained by accepting a
mobile. Thereafter, since $f_{AP}=0$ the policy rejects all incoming
mobiles due to decrease in network throughput and hence decrease in
corresponding reward, with increasing number of mobiles. Similarly,
for the dedicated mobiles to the NodeB network, the optimal policy
accepts new mobiles until the network throughput increases (Figure
\ref{theta_nodeb}) and rejects them thereafter due to absence of any
financial reward component and decrease in the network throughput.
Note that we have considered zero financial gains here
($f_{AP}=f_{3G}=0$) to be able to exhibit existence of these {\it
threshold type} policies for the dedicated streams.

\begin{figure}
\begin{minipage}{5in}
\centering \psfrag{ETA}{$\eta$} \psfrag{MUC}{$m_c$}
\psfrag{INN}{$\in$} \psfrag{TOO}{$\to$} \psfrag{PII}{$\pi$}
\psfrag{TAU}{$\tau$}
\includegraphics[width=2.9in]{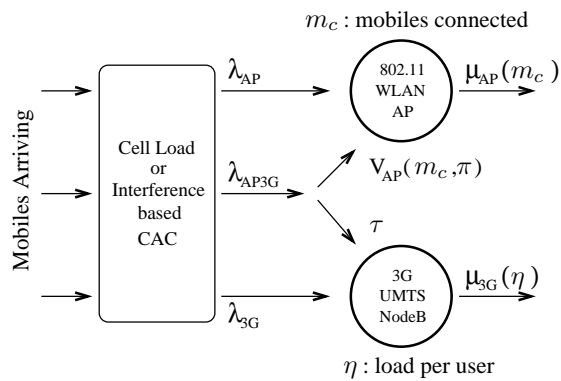}
\caption{Hybrid cell scenario under Individual optimality}
\label{hybrid-net-indiv}
\end{minipage}
\end{figure}

\section{Individual Optimality: Non-cooperative Dynamic Game }
\label{Individual-Optimality}

In the Individual Optimality approach here, we assume that an
arriving mobile must itself selfishly decide to join one of the two
networks such that its own cost is optimized. We consider the
{\it average service time} of a mobile as the decision cost criteria and
an incoming mobile connects to either the AP or NodeB network
depending on which of them offers minimum average service time. We
study this model within an extension of the framework of
\cite{EANS98} where an incoming user can either join a shared server
with a PS service mechanism or any of several dedicated servers.
Based on the estimate of its expected service time on either of the
two servers, a mobile takes a decision to join the
server on which its expected service time is least. This framework
can be applied to our hybrid cell scenario so that the AP is modeled
by the shared server and the dedicated DCH channels of the NodeB are
modeled by the dedicated servers. For simplicity, we refer to
the several dedicated servers in \cite{EANS98} as one single dedicated
server that consists of a pool of dedicated servers.
Then the NodeB comprising the dedicated DCH
channels is modeled by this single dedicated server and this type of framework
then fits well with our original setting in Section
\ref{mobile-arrivals}. Thus we again have an $M/G/2$ processing server
situation (see Figure \ref{hybrid-net-indiv}). As
mentioned before, the mobiles of dedicated streams directly join
their respective AP or NodeB network. Mobiles arriving in the common
stream decide to join one of the two networks based on their {\it
estimate} of the expected service time in each one of them. However,
an estimate of the expected service time of an arriving mobile $j$
must be made taking into account the effect of subsequently arriving
mobiles. But these subsequently arriving mobiles are themselves faced with a similar
decision problem and hence their decision will affect the
performance of mobile $j$ which is presently attempting to connect or
other mobiles already in service. This dependance thus induces a
non-cooperative game structure to the decision problem and we seek
here to study the Nash equilibrium solution of the game. The
existence, uniqueness and structure of the equilibrium point have
been proved in \cite{EANS98} already. Here we seek to analytically determine
the service time estimate and explicitly compute the equilibrium
threshold policy. As in \cite{EANS98}, a decision rule or policy for
a new mobile is a function $u : \{0,1,\ldots,M_{AP}-1\} \rightarrow
[0,1]$ where $M_{AP}$ is the pole capacity of the AP network. Thus
for each possible state of the AP network denoted by number of
mobiles already connected, $m_c$, a new mobile takes a randomized
decision $u(m_c) \in [0,1]$, that specifies the probability of
connecting to the AP. $1-u(m_c)$ then represents either the
probability of connecting to the NodeB or abandoning to seek a
connection altogether if both networks are full to their pole
capacity. A policy profile $\pi = (u_0,u_1,\ldots)$ is a collection
of decision rules followed by all arriving mobiles indexed
$(0,1,\ldots)$.

Define $V_{AP}(m_c,\pi)$ as the expected service time of a mobile in
the AP network, given that it joins that network, $m_c$ mobiles are
already present and all subsequent mobiles follow the policy profile
$\pi$. A single mobile generally achieves lower throughputs (i.e.,
higher service times) in a NodeB network as compared to in an AP
network. For simplification, we assume a worst case estimate for the
expected service time of a mobile in the NodeB network.
Denote ${\hat{\mu}}_{3G}:= \min_{\eta}{\mu_{3G}(\eta)}$ and
let $\tau:= 1/{\hat{\mu}}_{3G}$ be the maximum
service time of a mobile in the NodeB cell, which is independent of
network state $\eta$. For some $q$ ($0 \leq q \leq 1$, $q\in\mathbb{R}$),
define a decision policy $u(m_c)$ to be the
best response of a new mobile, against the policy profile
$\pi=(u_0,u_1,\ldots)$ followed by all subsequently arriving mobiles
\cite{EANS98}, as,
\[
u(m_c) = \left\{ \begin{array}{cc} 1 &\qquad :\; V_{AP}(m_c,\pi) < \tau  \\
        q &\qquad :\; V_{AP}(m_c,\pi) = \tau  \\
        0 &\qquad :\; V_{AP}(m_c,\pi) > \tau  \end{array}
\right
.
\]
Further, define a special kind of decision policy, namely the {\it threshold} policy
as, given $q$ and $L$ such that
$0 \leq q \leq 1$, $q\in\mathbb{R}$ and $L\geq 0$, $L\in\mathbb{Z}^+$, an
$L,q$ threshold policy $u_{L,q}$ is defined as,
\[
u_{L,q}(m_c) = \left\{ \begin{array}{cc} 1 &\qquad :\; m_c < L  \\
        q &\qquad :\; m_c = L  \\
        0 &\qquad :\; m_c > L \end{array}
\right
.
\]
This $L,q$ threshold policy will be denoted by $[L,q]$ or more compactly
by $[g]$ where $g=L+q$. Note that the threshold policies $[L,1]$ and $[L+1,0]$
are identical. We also use the notation ${[g]}^{\infty}\equiv {[L,q]}^{\infty}$
to denote the policy profile $\pi=([g],[g],\ldots)$.
Now, it has been proved in Lemma 3 in \cite{EANS98} that
the optimal best response decision policy $u(m_c)$ for a new mobile, against the policy profile $\pi$
followed by all subsequently arriving mobiles, is actually the threshold policy $[L^*,q^*]$
which can be computed as follows.
If $V_{AP}(M_{AP}-1,{[M_{AP}]}^{\infty}) < \tau$ then $L^*=M_{AP}$ and $q^*=0$.
Otherwise, let $L^{min} \defin \min \{L \in \mathbb{Z}^+ : V_{AP}(L,{[L,1]}^{\infty}) > \tau \}$.
Now, if $V_{AP}(L^{min},[L^{min},0]^\infty) \geq \tau$, then the threshold policy is given by
$[L^*,q^*] = [L^{min},0]$. Else if $V_{AP}(L^{min},[L^{min},0]^\infty) < \tau$ then it is given
by $[L^*,q^*]=[L^{min},q^*]$ where $q^*$ is the unique solution of the equation,
\begin{equation}
\label{qstar-equation}
V_{AP}(L^{min},[L^{min},q^*]^\infty) = \tau.
\end{equation}
Assuming state dependent service rate $\mu_{AP}(m_c)$ for a mobile
in the AP network, we now compute $V_{AP}(m_c,\pi)$ analytically. At
this point we would like to mention that the derivation of the
entity equivalent to $V_{AP}(m_c,\pi)$ in \cite{EANS98} is actually
erroneous. Moreover the basic framework in \cite{EANS98} differs
from ours, since in our framework we have dedicated arrivals in
addition to the common arrivals and we consider a {\it state
dependent} service rate $\mu_{AP}(m_c)$ for the shared AP server.
For notational convenience, if $V(m_c) \defin
V_{AP}(m_c,[L,q]^\infty), 0\leq m_c \leq M_{AP}-1$, then it is the
solution of the following set of $M_{AP}$ linear equations, where
$\alpha := \lambda_{AP} + \lambda_{AP3G} + \mu_{AP}(m_c)$
(dependence of $\alpha$ on $m_c$ has been suppressed in the
notation):

Case 1: $4 \leq L \leq M_{AP}-2$,
\begin{equation}
\begin{split}
V(0) & = \frac{1}{\alpha} + \frac{\lambda_{AP} + \lambda_{AP3G}}{\alpha} V(1) \\
V(m_c) & = \frac{1}{\alpha} + \frac{\mu_{AP}(m_c)}{\alpha}\frac{m_c}{m_c+1}V(m_c-1) 
               + \frac{\lambda_{AP} + \lambda_{AP3G}}{\alpha} V(m_c+1), \\
          & \hspace{9cm} 1\leq m_c \leq L-2 \\
V(L-1)& = \frac{1}{\alpha} + \frac{\mu_{AP}(L-1)}{\alpha}\frac{L-1}{L}V(L-2) 
        + \frac{\lambda_{AP} + q \; \lambda_{AP3G}}{\alpha}V(L) \\
	& \quad + \frac{\lambda_{AP3G}}{\alpha}(1-q)V(L-1) \\
V(L) & = \frac{1}{\lambda_{AP}+\mu_{AP}(L)} + \frac{\mu_{AP}(L)}{\lambda_{AP}+\mu_{AP}(L)}\frac{L}{L+1} V(L-1) 
       + \frac{\lambda_{AP}}{\lambda_{AP}+\mu_{AP}(L)} V(L+1) \\
V(m_c) & = \frac{1}{\lambda_{AP}+\mu_{AP}(m_c)} + \frac{\mu_{AP}(m_c)}{\lambda_{AP} + \mu_{AP}(m_c)}\frac{m_c}{m_c+1} \\
        & \quad \times V(m_c-1) + \frac{\lambda_{AP}}{\lambda_{AP} + \mu_{AP}(m_c)} V(m_c+1), \quad L+1\leq m_c \leq M_{AP}-2 \\
V(M_{AP}-1) & = \frac{1}{\mu_{AP}(M_{AP}-1)} + \frac{M_{AP}-1}{M_{AP}} V(M_{AP}-2)
\end{split}
\end{equation}
Case 2: $L=M_{AP}-1$,
\begin{equation}
\begin{split}
V(0) & = \frac{1}{\alpha} + \frac{\lambda_{AP} + \lambda_{AP3G}}{\alpha} V(1) \\
V(m_c) & = \frac{1}{\alpha} + \frac{\mu_{AP}(m_c)}{\alpha}\frac{m_c}{m_c+1}V(m_c-1) 
               + \frac{\lambda_{AP} + \lambda_{AP3G}}{\alpha} V(m_c+1), \quad 1\leq m_c \leq L-2 \\
V(L-1) & = \frac{1}{\alpha} + \frac{\mu_{AP}(L-1)}{\alpha}\frac{L-1}{L}V(L-2) + \frac{\lambda_{AP} + q \; \lambda_{AP3G}}{\alpha}V(L) \\
	       & \quad + \frac{\lambda_{AP3G}}{\alpha}(1-q)V(L-1)\\
V(L) & = \frac{1}{\mu_{AP}(L)} + \frac{L}{L+1} V(L-1).
\end{split}
\end{equation}
%

\begin{figure}
\begin{minipage}{5in}
\centering
\includegraphics[width=3in]{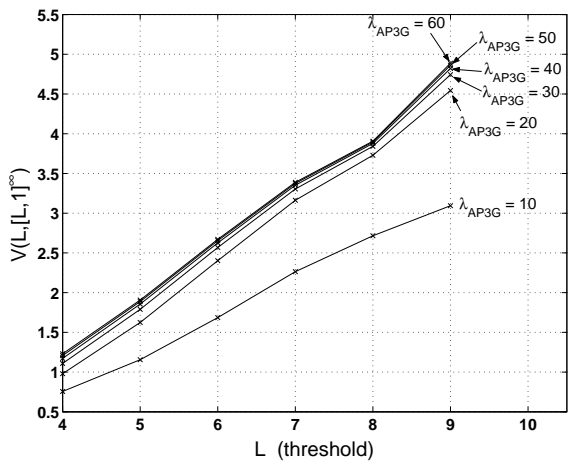}
\caption{$V_{AP}(L,[L,1]^\infty)$ v/s $L$ for
$\lambda_{AP}=3$ and $M_{AP}=10$} 
\label{V_L_L_1}
\end{minipage}
\end{figure}

\begin{figure}
\begin{minipage}{5in}
\centering
\includegraphics[width=3in]{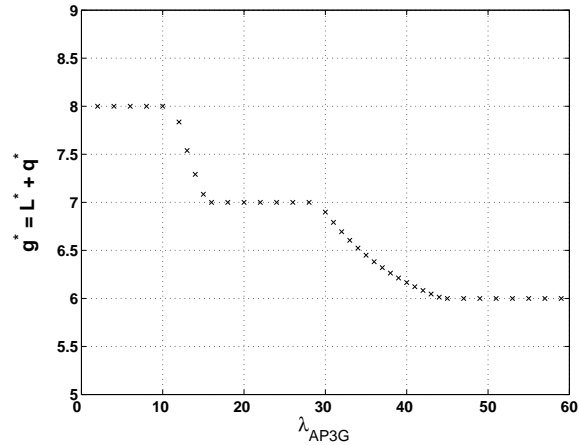}
\caption{$g^*$ v/s $\lambda_{AP3G}$ for
$\lambda_{AP}=3$, $M_{AP}=10$ and $\tau=2.5$} \label{gstar}
\end{minipage}
\end{figure}

The above system of $M_{AP}$ linear equations with $m_c=L$ and $q=1$
can be solved to obtain
$V_{AP}(L,[L,1]^\infty)$ for different values of $L$. Figure \ref{V_L_L_1}
shows an example plot for $\zeta=10^{-5}$, $\lambda_{AP}=3$, $M_{AP}=10$,
$M_{3G}=10$ and other numerical values for various entities in WLAN and UMTS
networks being the same as those used in Section \ref{semi-mdp-control}.
Assuming a certain pole capacity $M_{3G}$ of the NodeB cell, $\tau$ can be
computed from its definition and Equation \ref{UMTS_pole_cap}. Knowing $\tau$,
one can compute $L^{min}$ from Figure \ref{V_L_L_1} and then finally $q^*$
from Equation \ref{qstar-equation}. Figure~\ref{gstar} shows a plot of the
equilibrium threshold $g^*=L^*+q^*$ against $\lambda_{AP3G}$, with computed
value of $\tau=2.5$ for $M_{3G}=10$ and $\lambda_{AP}=3$. As in \cite{EANS98}, the
equilibrium threshold has a special structure of {\it descending staircase}
with increasing arrival rate ($\lambda_{AP3G}$) of mobiles in common stream.


\section{Conclusion}
\label{conclusion} In this paper, we have considered optimal
user-network association or load balancing in an AP-NodeB hybrid
cell. We have studied two different and alternate approaches of
Global and Individual optimality under SMDP decision control and
non-cooperative dynamic game frameworks, respectively. To the best
of our knowledge, this study is the first of its kind. Under global
optimality, the optimal policy for common stream of mobiles has a
neither convex nor concave type switching curve structure, where as
for the dedicated streams it has a threshold structure. Besides, a
{\it mobile-balancing} and a {\it mobile-greedy} network phenomenon
is observed for the common stream. For the analogous AP-AP
homogenous network setup, a threshold type and symmetric switching
curve is observed. An interesting switching curve is obtained for
the NodeB-NodeB homogenous case, where the state space is divided
into L-shaped and quadrilateral shaped regions. The optimal policy
under individual optimality model is also observed to be of
threshold type, with the threshold curve decreasing in a staircase
fashion when plotted against increasing arrival rate of the mobiles
of common stream.

\end{document}